\documentclass{emulateapj}
\begin{document}

\title{PROTO-BROWN DWARF DISKS AS PRODUCTS OF PROTOSTELLAR DISK ENCOUNTERS}
\author{Sijing Shen and James Wadsley}
\affil{Department of Physics and Astronomy, McMaster University}
\affil{Hamilton, Ontario L8S 4M1, Canada}  
\email{shens@physics.mcmaster.ca}

\begin{abstract}

  The formation of brown dwarfs via encounters between proto-stars has
  been confirmed with high-resolution numerical simulations with a
  restricted treatment of the thermal conditions.  The new results
  indicate that young brown dwarfs (BDs) formed this way are disk-like
  and often reside in multiple systems.  The newly-formed proto-BDs
  disks are up to 18 AU in size and spin rapidly making small-scale
  bipolar outflows, fragmentation and the possible formation of
  planetary companions likely as have recently been observed for BDs.
  The object masses range from 2 to 73 Jupiter masses, distributed in
  a manner consistent with the observed sub-stellar initial mass
  function.  The simulations usually form multiple BDs on eccentric
  orbits about a star.  One such system was hierarchical, a BD binary
  in orbit around a star, which may explain recently observed
  hierarchical systems.  One third of the BDs were unbound after a few
  thousand years and interactions among orbiting BDs may eject more or
  add to the number of binaries.  Improvements over prior work include
  resolution down to a Jupiter mass, self-consistent models of the
  vertical structure of the initial disks and careful attention to
  avoid artificial fragmentation.

\end{abstract}


\section{INTRODUCTION}

Brown dwarfs (BDs) are abundant, both in the field and in star
clusters \citep{chabrier,martin,luhman}. The flat low mass end of
observed stellar initial mass function (IMF) \citep[e.g.,][]{hambly,
  martin_98,barrado} suggests that brown dwarfs are as numerous as
hydrogen-burning stars.  Observations indicate that a large percentage
of young BDs are surrounded by circumstellar disks \citep{muench}.
Mid-infrared, submillimetre and millimetre measurements have
characterized the spectral energy distributions (SEDs) of the BD disks
\citep{pascucci}, revealing the accretion rate and ongoing evolution
of the young BD disks.  The decreasing trend of IR excess in SEDs with
age suggests that BD disks evolve from flared to flat geometry,
similar to the evolution of disks surrounding large stars
\citep{mohanty}.  A T-Tauri phase has been detected via broad
$H_{\alpha}$ lines in many of these young disks \citep{muzerolle,
  jaya}), sometimes associated with bipolar outflows \citep{whelan}.
Giant planets have also been imaged orbiting brown dwarfs
\citep{chauvin}.

While the observational results admit the possibility that brown
dwarfs share a common origin with stars (i.e. from the gravitational
collapse of molecular cloud cores), the theory of the BD formation is
far from well established.  BD masses ($ 0.013-0.075 \ {\rm
  M}_{\odot}$) are smaller than the typical Jeans mass in the parent
clouds ($\sim 1 \ {\rm M}_{\odot}$) by more than an order of magnitude
which seems to work against direct collapse of clumps as a formation
mechanism.  Approaches to solving this problem include lowering the
Jeans Mass via regions of low temperature \citep{elmegreen} or via
density enhancements in turbulent flows \citep{padoan} (``turbulent
fragmentation'').  After low mass objects form ongoing mass accretion
determines the final IMF.  It has been proposed that all objects
initially form with low masses and that observed BDs are those objects
ejected from the gas cloud reservoir before they grow to stellar mass
(``competitive accretion'') \citep{cluster_bate, reipurth}.

Here we investigate a complementary mechanism of brown dwarf
formation: instead of producing sub-stellar object in large numbers
directly, we start by using the fact that initial cloud core-collapse
stage forms numerous stellar mass objects with extended protostellar
disks.  Such early disks are very flared and for this study they are
assumed to be stable against spontaneous fragmentation.  However,
encounters between extended disks are very violent and can trigger
gravitational instabilities and the formation of
substellar objects.  Since the gas density in a disk is much higher
than an early-stage core, the local Jeans mass is comparable to
a substellar mass, making it possible for gravitational collapse to
produce objects that range down to planetary masses.

The encounter picture has been studied previously by \citet{watkins,
  watkinsb} and \citet{lin} with low resolution simulations using
proto-stars with flat disks (two dimensional) or disks with
highly-simplified vertical structure, respectively.  The formation of
substellar-mass objects was reported by both groups but it was not
clear if the fragmentation was real or numerical or what the detailed
properties of the objects might be.  It is also worth noting that
encounters occur in gaseous simulations of star cluster formation
(e.g. \cite{cluster_bate}).  However, modelling many hundreds of solar
masses at once forces a compromise on the the mass resolution in
individual proto-stellar disks and approximations (such as unusually
low mass clusters or very large sink radii, discussed in the next
section) have been necessary.  

In this letter we report the results of new, high-resolution pairwise
encounter simulations with realistic protostellar disk models and a
range of initial configurations.  We discuss the conditions under
which fragmentation can occur and, in the cases where objects form, we
examine their shapes, rotation and orbital properties and discuss
their observational implications.

\section{SIMULATIONS}

\begin{figure}
\plotone{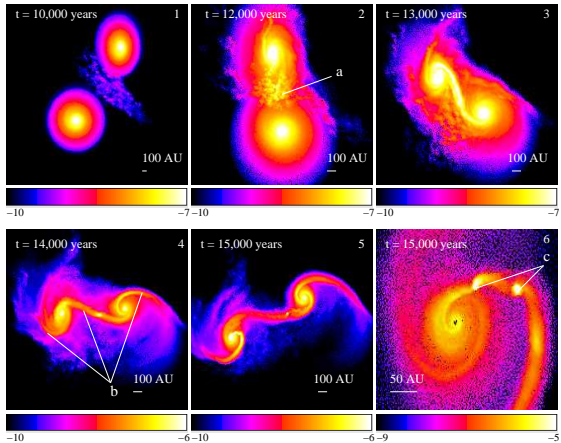}
\caption{An encounter between two prograde rotating protostellar
  disks.  The disk moving from top-right to bottom-left was tilted
  $\pi/4$ relative to the encounter plane.  Panels 1-5 correspond to
  times 10000, 12000, 13000, 14000, and 15000 years, respectively. Panel
 6 shows the fragmented disk at bottom-left of panel 5 with different
 zoom and density scale. Shock compressed gas (a), tidal structures (b) and
 strong disk instabilities (c) are indicated. The color bars indicate the
 density scale in logarithmic space in units of
 ${\rm M}_{\odot}/{\rm AU}^{3}$. Note that the black regions in panels 1-5 are
 filled with gas,  but have density $\lesssim 10^{-10} \  {\rm
   M}_{\odot}/{\rm AU}^{3}$.
 At 15000 years, BD Clumps are present in the central tidal feature and the inner disk
 (c) in the last panel.}
\label{fg:encounter}
\end{figure}

The parallel TreeSPH code ``Gasoline'' \citep{wadsley} was used to
simulate the encounters of protostellar disks.  The initial conditions
employed flared disk models \citep{chiang} consistent with the
observations of early stellar objects.  Each disk extended from 50 to
$\sim 1500$ AU and had mass $\sim 0.6 \ {\rm M}_{\odot}$, comparable
to the mass of the central protostar ($\sim 0.5 \ {\rm M}_{\odot}$).
The radial surface density and temperature profiles were chosen to be
similar to the intermediate quasi-static results of core-collapse
simulations by \citet{yorke}.  The dependence is flat inside 100 AU
and then varies as $r^{-3/2}$ until it falls off rapidly beyond 500-700 AU.

The gas temperature was fixed according to the distance to the nearest
protostar as $T(r) \propto r^{-1/2}$ \citep{mayer, pickett, chiang}
and smoothly transitioned to the ambient temperature of $\sim 40 $ K
at $r= 500$ AU.  This temperature profile is similar to that of
\cite{yorke} which self-consistently included heating due to stellar
irradiation, viscosity and shock waves which are beyond current
simulation capabilities for non-axisymmetric simulations.  Based on
Yorke's surface density and using the \cite{dalessio} opacities the
disks are optically thin outside clumps and cool efficiently.  The
product of the angular rotation rate and the cooling time $\Omega\
t_{cool}$ = 0.3-0.7 which is well within the required range for
fragmentation for a molecular hydrogen disk of $\Omega\ t_{cool} < 12$
\citep{rice, gammie}.  Given the importance of heating and the
extremely short cooling time ($t_{cool}\sim 30$ years at 100 AU) it
was felt that a fixed temperature distribution was most appropriate
for the protostellar disks.  Better treatment of the thermal
conditions will be required for interior evolution of the brown dwarf
clumps being pursued in future work.

Vertical structure is one of the key factors affecting disk stability,
lowering the critical Toomre $Q$ value to values as low as $\sim 0.6$
for extremely flared disks \citep{kim}.  Following arguments by
\cite{yorke} and \cite{mayer} and experimentation with different $Q$
values \citep[][in preparation]{shen}, we selected an initial $Q_{MIN}
= 1.6$ that is comfortably in the stable region for the simulations
presented here.  The disk vertical structure was constructed
self-consistently assuming the gas to be in hydrostatic equilibrium
with both its own global self-gravity and the star gravity.  The scale
heights in the midplane are quite large ($\sim 6$ AU at the densest
point at $\sim 80-100$ AU) and well resolved (local SPH smoothing 1.9
AU) and decrease above the mid-plane.  The velocities were set to
maintain circular orbits.

Each disk was modeled with 200,000 SPH particles with masses of $3
\times 10^{-6} \ {\rm M}_{\odot}$ (1 Earth Mass).  An unresolved Jeans
mass can result in artificial fragmentation \citep{res_bate}.  This
resolution ensured that both the local Jeans mass and the vertical
structure were resolved until well after any initial fragmentation
that forms clumps.  The disks were tested to be initially stable
against spontaneous fragmentation.  The artificial shock-capturing
viscosity had no dynamical effect (measured viscous timescales of
$10^{8-11}$ years).  The encounter timescale ($\sim 10^4$ years) is
too short for physical viscous effects to play an important role.

The protostars the stars were treated as sink particles
\citep{cluster_bate} with radii of 1 AU but no new sink particles were
added during the runs.  The softening length of gravitational force
was chosen to be 0.2 AU, one tenth of the minimum initial spacing.
Thus densities up to 1000 times larger than the highest densities in
the initial disk were well treated numerically.  The softening was
expected to inhibit extremely low-mass fragmentation near the
resolution limit.

Pairs of disks were placed on initially unbound orbital
trajectories with a projected impact parameter of $1000$ AU (prior to
gravitational focusing).  Various disk spin orientations (with respect
to the angular momentum of the orbit) were run, covering prograde
cases (spin-orbit angle less than $\pi/2$), retrograde cases
(spin-orbit angle larger than $\pi/2$) and combinations of the two.

\section{RESULTS AND DISCUSSION}

\begin{figure}
\plotone{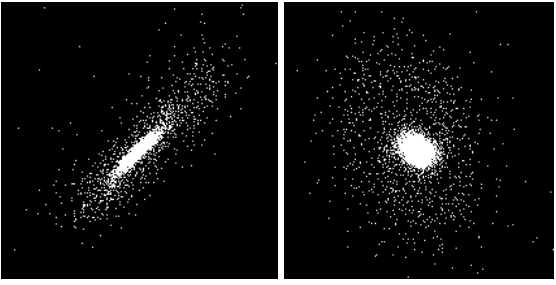}
\caption{Typical shape of newly formed Brown Dwarf mass objects. The specific object shown was
  produced via shock layer fragmentation in a coplanar disk
  encounter between disks rotating in a retrograde sense. The encounter occurred
  in the x-y plane. \emph{left:} the object projected in the
  x-y plane; \emph{right:} the same object projected in x-z plane.
  Both boxes are $3 \ {\rm AU} \times 3 \ {\rm AU}$ across.}
\label{fg:fragspin}
\end{figure}
 
The simulation results support the main conclusion of the previous
studies: the formation of BDs can be triggered by protostellar disk
encounters.  However, the conditions for fragmentation are more
restrictive than previously reported, such that previously employed
initial conditions (e.g. $Q_{MIN}=2.4$) are unable to fragment when fully
resolved.  Vertically resolved disks closer to the critical value of
stability criterion are required ($Q_{MIN}\leq 2.1$).  Convergence studies
also indicate that artificial fragmentation is likely with resolution
significantly less than that employed for our production runs
\citep[][in preparation]{shen}.  The primary new results, reported
here, are the diskiness of the young BD, the extension of the mass
distribution down to planetary masses and the likelihood of multiple
or even hierachical systems.

A typical prograde encounter is depicted in Fig~\ref{fg:encounter}.
Because of the tidal interaction, gas in the outer part ($R > 300-500$
AU) disperses.  Fragmentation usually occurred after the disks
passed periastron. In the case depicted here, a 20 $ M_{\rm
  {Jupiter}}$ clump formed in a distant tidal arm structure at t = 14,000 years and
two objects (with mass 9 and 70 $ M_{\rm {Jupiter}}$) formed via
gravitational instability in an inner disk at t = 15,000 years.  Three
basic mechanisms were found for forming sub-stellar mass objects:
fragmentation within shock layers, condensations within prograde tidal
structures and fragmentation via instabilities in the inner disks
after inflows.  Shocks (Fig. \ref{fg:encounter}, label a)
formed due to the highly supersonic velocities of the impacts, but
fragmentation in shocks only happened in coplanar encounters.  Tidal
tails (Fig. \ref{fg:encounter}, label b) were induced by gravitational
resonance in prograde disks.  Disk instability (Fig.
\ref{fg:encounter}, label c) was a common feature during encounters,
as mass moved inwards due to angular momentum transport via tidal
tails for prograde disks or cancellation of rotation in retrograde
cases.

The self-gravitating objects produced in the simulations are in the
mass range from 2 (600 particles) to 73 Jupiter masses, covering the
BD and giant planet regimes.  All of the objects are highly flattened
and disk-like (Fig. \ref{fg:fragspin}).  A T-Tauri-like phase,
implying a disk, has been observed for planetary mass object
\citep{mohanty_2005}.  At the end of the simulations, the proto-BD
disks had radii 0.3-18 AU, with the lower limit quite likely related
to the gravitational softening (0.2 AU) and height-to-radius ratios of
$H/R \sim 0.1 $.  The objects spin rapidly and most of the clumps
formed via fragmentation of protostellar disks have similar spin
directions to the parent disk.  But it is not necessary true for the
clumps formed in shock layers or tidal tails.

The simulations were halted several thousand years after BD-mass
objects form when the stars are well past peri-astron.  During further
evolution, gas in the BD disks with lower specific angular momentum
should condense to form a central core, and the remaining disk will
accrete onto the core on a viscous time scale.  In order to for these
fast-rotating disks lose their angular momentum, it is natural to
expect observable bipolar outflows.  These were confirmed
observationally via the recent detection of several forbidden lines
({\it e.g.} $[O_{I}] \& [S_{II}]$) and a P Cygni-like dip in the
$H_{\alpha}$ profile in the spectrum of a young BD $\rho$ Oph 102
\citep{whelan}.  The lower accretion rate expected in a BD disk
\citep{natta} makes it a scaled-down version of a T Tauri star as
shown by the observed offset from the continuum of the forbidden
emission lines in a BD-outflow \citep{whelan}.  Another way to
distribute the angular momentum of the proto-BD disk is to form
planets or further fragment.  Thus BD companions can have planetary
masses, forming a BD-planet system, or comparable masses to the
primary BD to form a binary or multiple BD system.  The giant planet
companion (2M1207 b) to a nearby young BD found by the new VLT/NACO
imaging observations \citep{chauvin} is a possible example of the
former case.  The large inferred inner holes in the spectra of some BD
disks have been considered as signs of ongoing planet formation
\citep{mohanty}.

Formation of binary BD pair occurred in one of our simulations via the
secondary fragmentation of a large (6 AU) clump $\sim 140$ AU from the
star.  Though with 5000 particles the clump is larger than the the
entire protostellar disk in previous work, the Jeans Mass is only
marginally resolved and this result should be confirmed with more
resolution and additional physics.  We note that fragmentation is not
unexpected from a physical perspective given the high rate of spin
\cite{matsumoto}.  The two components (19 and 13 Jupiter masses)
orbited each other with separation $\sim$ 3 AU and orbital period
$\sim$ 30 years.  Both retained a highly-flattened shape.
It is likely that the system will remain as an hierarchical star-BD
multiple.  This case is observationally inferred for GL 569 where the
primary is a M2.5V star (GL 569A) and a recently
observationally-confirmed BD triple (GL 569B) \citep{gorlova,simon}.
The age of the quadruple is about 100-125 Myr so no disks are
observed.  However, in some of much younger BD multiples, accretion
signatures are indicated via a V-band excess \citep{kraus}, suggesting
accretion disks around the components.

\begin{figure}
\plotone{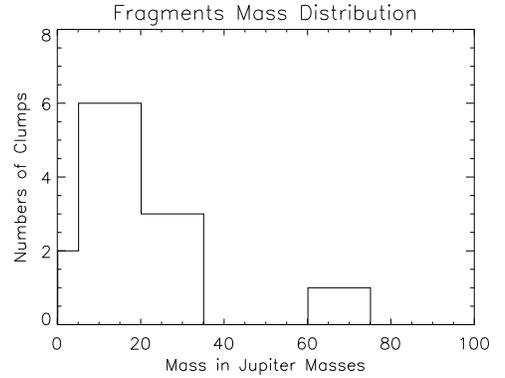}
\caption{A preliminary mass distribution of 32 objects resulting from 14
  simulations.  All proto-stellar disks in these simulations had similar vertical 
  structure and were all stable against spontaneous
  fragmentation.  The encounter velocities, the relative angle between
  two disks and the angle between the disk and the encounter plane
  differed from case to case.}
\label{fg:massdist}
\end{figure}

The rapid growth assumption of the competitive accretion scenario
\citep{reipurth, cluster_bate} makes ejection a requirement in order
to retain BDs.  In our simulations, about one third of the clumps (12
out of 32) were unbound and left the parent disks with 1D RMS velocity
dispersion $v \sim 2.6 \ {\rm km/s}$, higher than the velocities of
the protostars.  Among the bound BDs, three out twenty have large
semi-major axes ($a > 300 $ AU).  As gas dispersion is severe in the
outer disk region due to the interaction, these objects will lose
access to a gas reservoir and become free-floating BDs or
wide-separation BD companions as seen in observations
\citep[e.g.,][]{wide_bd}.  For objects close to the star ($a < 100$
AU), ejection may occur due to interactions with other orbiters.  The
need to invoke ejection to shut off rapid growth is still under
debate. For example, \citet{krumholz} suggest that accretion is very
slow and the final mass of an object is determined by the initial
collapse and thus BD-mass objects should remain as brown dwarfs.  The
competitive accretion model also needs to better establish that the
first objects to form in a cluster environment will have such low
masses.

Encounter-induced protostellar disk fragmentation provides an
additional mechanism to produce brown dwarfs and it does not rule out
BD formation via other methods such as turbulent fragmentation.
Ongoing work involving more encounter cases will build up statistics
to enable estimates of which of these models is more likely to produce
observed sub-stellar populations based on the production rates, BD
disk properties, the frequency of multiples and the impact on
proto-stellar evolution and accretion histories.

A preliminary mass distribution of the simulated young BDs is plotted
in Fig. \ref{fg:massdist}.  The clumps were found to be more abundant
at the lower mass end, which is broadly consistent with observed
substellar initial mass function \citep{martin, barrado}. The turnover
at the lowest (planetary) mass ($M < 10 \ M_{\rm {Jupiter}}$) is
related to our resolution limits.  An apparent gap in the population
from 35-60 $ M_{\rm {Jupiter}}$ is suggested in the figure, but the
current statistics are insufficient to confirm it and additional runs
will be required.  Clumps typically form more than 100 AU from a star
but sometimes significantly closer where the physical conditions are
more complex.  However, there is no apparent bias (according to radius
of formation) in the mass distribution and it looks similar if clumps
within 100 AU are excluded.

If encounter-induced BD formation is dominant, the BD numbers
(relative to the number of stars) is expected to differ depending on
the star-forming environment.  Observation indications of this have
been reported \citep[e.g.,][]{levine, luhman_2003, briceno} but the
data are not conclusive.  The way this ratio varies as a function of
cluster properties is not straightforward.  For example, an increase
in the cluster density indicates more impacts between its members,
while a higher velocity dispersion in denser cluster greatly reduces
the number of fragments per encounter \citep[][in preparation]{shen}.
To make a concrete prediction regarding BD production as a function of
star cluster properties, the more comprehensive set of simulations now
underway will have to be combined with detailed estimates of encounter
rates in young star clusters.

\end{document}